# Strategies to enhance THz harmonic generation combining multilayered, gated, and metamaterial-based architectures

Ali Maleki,[1] Moritz B. Heindl,[2] Yongbao Xin,[3] Robert W. Boyd,[1,4,5] Georg Herink,[2] Jean-Michel Ménard[1,4,*]

[1]*Department of Physics, University of Ottawa, Ottawa, ON K1N 6N5, Canada*
[2]*Experimental Physics VIII – Ultrafast Dynamics, University of Bayreuth, Bayreuth, 95447, Germany*
[3]*Iridian Spectral Technologies Ltd, Ottawa, ON K1G 6R8, Canada*
[4]*School of Electrical Engineering and Computer Science, University of Ottawa, Ottawa, ON K1N 6N5, Canada*
[5]*Institute of Optics and Department of Physics and Astronomy, University of Rochester, NY 14627, USA*

* Corresponding autour: Jean-Michel Ménard (jean-michel.menard@uottawa.ca), Tel :+1 613-562-5800 ext.2318

## Abstract

Graphene has unique properties paving the way for groundbreaking future applications. Its large optical nonlinearity and ease of integration in devices notably makes it an ideal candidate to become a key component for all-optical switching and frequency conversion applications. In the terahertz (THz) region, various approaches have been independently demonstrated to optimize the nonlinear effects in graphene, addressing a critical limitation arising from the atomically thin interaction length. Here, we demonstrate sample architectures that combine strategies to enhance THz nonlinearities in graphene-based structures. We achieve this by increasing the interaction length through a multilayered design, controlling carrier density with an electrical gate, and modulating the THz field spatial distribution with a metallic metasurface substrate. Our study specifically investigates third harmonic generation (THG) using a table-top high-field THz source. We measure THG enhancement factors exceeding thirty and propose architectures capable of achieving a two-order-of-magnitude increase. These findings highlight the potential of engineered graphene-based samples in advancing THz frequency conversion technologies for signal processing and wireless communication applications.

## Introduction

Nonlinear optics in the terahertz (THz) region has emerged as a promising field with diverse scientific and technological applications, fostering innovations in optical devices, advancements in material analysis, and imaging[1–3]. Intense THz fields interacting with nonlinear materials offer pathways for fundamental insights, from studying solid-state materials to capturing carrier and phonon dynamics[4–6]. To enable ultrahigh-speed information and communication technologies, many efforts are also invested to achieve efficient nonlinear THz frequency converters[7]. An archetypal illustration of this process is high harmonic generation (HHG), where resulting photons are generated at energies corresponding to multiples of the incident photons' energy[8].

Recent years have witnessed a broadening landscape of efficient THz HHG platforms relying on various materials. They include superconductors[9,10], transition metal oxides (like $SrTiO_3$)[11], and most notably, Dirac fermion systems. Dirac electronic band structures exhibit a linear energy-momentum dispersion that has been associated with a strong nonlinear response at THz frequencies. These effects have been studied in different material geometries, such as one-dimensional (1D) massless Dirac systems (e.g. carbon nanotubes)[12], 2D semimetals and topological insulators (e.g. $Bi_2Se_3$)[13–15], and 3D Dirac semimetals (e.g. $Cd_3As_2$)[16]. HHG in 2D Dirac-fermion graphene[17] has been reported across a wide spectral range, from the near-infrared[18], to the mid-infrared[19–21] and the THz[22–24] regions. In general, the nonlinearity in graphene is attributed to both intraband and interband electron dynamics, taking advantage of its atomic thinness to prevent dephasing and dispersion of propagating waves[22]. In the THz region, previous work on graphene relying on accelerator-based super-radiant THz sources reported a strong nonlinear response ($\chi^{(3)} \sim 10^{-9}\ m^2/V^2$) resulting in an efficient HHGs conversion efficiency of 0.1%[23]. This phenomenon is attributed to the collective oscillation of massless Dirac fermions through intraband transitions[22,23,25], highlighting the unique characteristics and potential applications of graphene in this frequency range. Remarkably, this nonlinear response in graphene is achievable at moderate THz pump field strengths ranging between 10 to 90 kV/cm, at room temperature, and under ambient conditions. To enhance nonlinear effects in graphene-based samples in the THz region, several approaches have been used. For example, previous work has shown that an optimization of the carrier density in graphene can enhance the power conversion efficiency of the THz THG[26]. Metasurfaces can also be used to locally enhance an incident pump field to enable stronger HHG in a graphene sheet, reaching up to a 1% field conversion efficiency at an incident field strengths <30 kV/cm[27]. These more efficient approaches to achieve nonlinear effects have been demonstrated independently in the context of a single graphene layer only. Yet, the nonlinear optical interaction length inside two-dimensional (2D) materials only spans over one atomic layer, thus drastically limiting any frequency conversion processes. Therefore, stacking multiple graphene layers appears as a natural route towards the demonstration of more efficient graphene-based nonlinear devices. In this context, however, it remains unclear if the variation of carrier concentration and the use of a metasurface substrate can further enhance the nonlinear effects.

Here, a table-top high-field THz source is used to investigate the enhancement of third harmonic generation (THG) in chemical vapor deposition (CVD) graphene through three distinctive approaches. We first investigate the nonlinear response of stacked decoupled graphene sheets, ranging from 1 to 15 layers. Results show a correlation between the field strength of the generated third harmonic and the increasing number of graphene layers with a maximum nonlinear signal observed for 6 layers, partially due to a trade-off between enhanced interaction length and linear absorption. Then, we use an electrical gate to investigate the effect of doping concentration in a 1-, 2-, and 3-layer graphene stack. Finally, we compare three types of metallic metasurface acting on the THz driving field as a bandpass filter, a bandstop filter, and a linear polarizer. A simple model is proposed to estimate the THG enhancement factor based on the metasurface geometry. This study on a various set of graphene-based architectures provides privileged insight on the combination of distinct experimental strategies to increase frequency conversion processes in 2D materials.

## Results

### Stacked graphene layers

We explore THG in CVD graphene within the THz region using a table-top time-resolved THz setup generating single-cycle pulses from a nonlinear lithium niobate crystal[28] with a peak field of 360 kV/cm (see Methods). Multilayered graphene samples are fabricated on a THz transparent Zeonor substrate by successively depositing single graphene layers onto each other with the wet-transfer method (see Methods) while a 60 nm-thick aluminum oxide ($A_2lO_3$) spacing layers between graphene sheets reduces interlayer interactions. The total sample thickness remains significantly sub-wavelength, eliminating the need to satisfy phase matching conditions to achieve coherent nonlinear effects.

The experimental configuration is shown schematically in Fig. 1. Sensitive detection of the THG signal requires the use of a lowpass filter (LPF) reducing the spectral width of the pump pulse and eliminating any residual pump at the third harmonic frequency. After the generation of the THG inside the graphene-based sample, a highpass filter (HPF) is used to limit the incident pump reaching the detection scheme. This technique significantly increases the sensitivity of time-resolved detection by reducing the overall level of noise[23,27]. The general design of these filters are described in previous work[29] and their spectral transmission properties are shown in Section 2 of the Supplementary Information. In brief, the LPF transmits multicycle THz pump pulse centered at the frequency, $\omega = 0.8$ THz, and strongly attenuates (~50 dB) spectral components beyond 1.5 THz. The HPF attenuates the pump pulse by >30 dB while transmitting the third harmonic at $3\omega$.

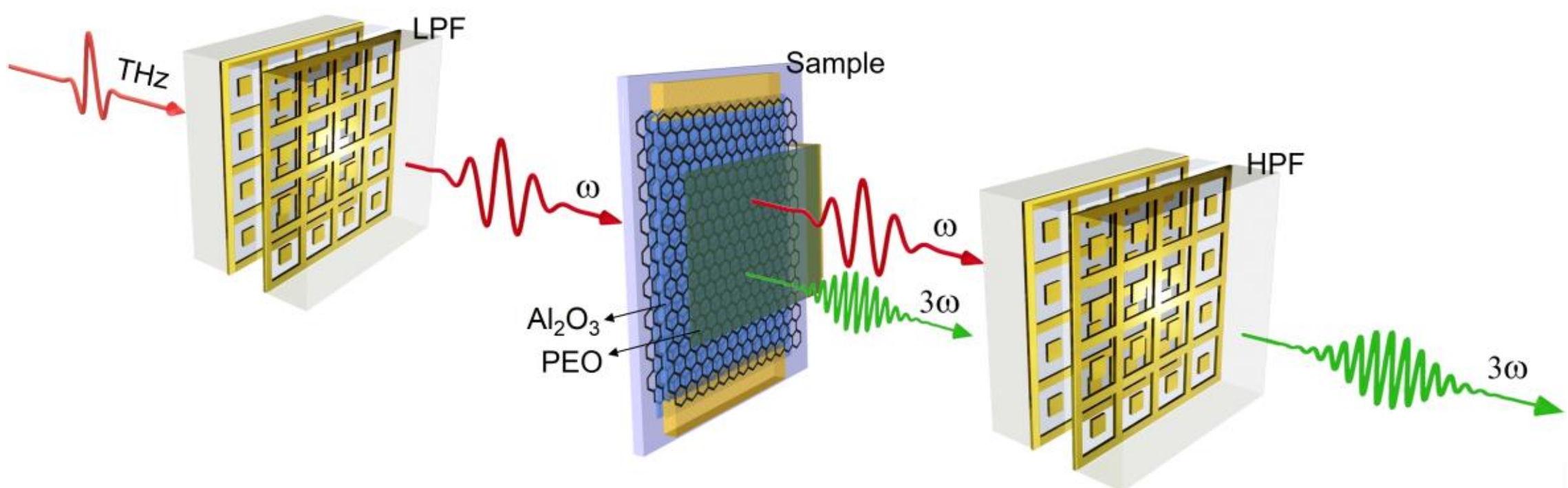

**Fig. 1 Schematic of the experimental configuration at the sample position to generate and detect THz third harmonic generation.** A LPF transmits a multicycle fundamental pulse at $\omega$, which produces a signal at $3\omega$ after nonlinear interactions in a graphene-based sample. The LPF and HPF consist of a superposition of four metasurfaces, but two only are schematically shown here for clarity. The graphene sheet is attached to electrodes (top and bottom), functioning as source and drain terminals for electrical characterization. Gating is facilitated by a polymer electrolyte [LiClO4:polyethylene oxide (PEO)] layer deposited on top of the graphene sample.

We investigate six nonlinear samples consisting of stacked graphene sheets, from 1 to 15 layers, allowing us to vary the nonlinear interaction length. Fig. 2a shows the spectrum of the THz pulse (blue line) transmitted through the LPF and used as the pump at the fundamental frequency $\omega$. This graph only contains spectral measurements collected with the 1-, 3-, and 6-layer graphene samples (purple, green and red lines, respectively) for enhanced clarity of the display. A distinct THG spectral peak at $3\omega = 2.4$ THz can be observed with an intensity increasing roughly quadratically with the number of graphene layers. The HPF, which role is to decrease the residual pump, allows the noise level at frequencies beyond 1.5 THz to settle close to the noise floor (grey shaded area), enabling sensitive monitoring of the nonlinear effects. Fig. 2b shows the time-resolved multicycle pump pulse (blue line) with a peak amplitude of 32 kV/cm. This signal is compared to the THG waveform (red line) generated by a 6-layer graphene sample. The time-resolved pump pulse is measured directly by electro-optic sampling in the experiment. The third harmonic waveform, weak in amplitude in comparison to the residual pump, is extracted by numerically applying a spectral bandpass filter around 2.4 THz and then using the inverse Fourier transform. Fig. 2c shows the third harmonic peak amplitude obtained with all the samples investigated in this experiment. The largest nonlinear signal is obtained with the stack of 6 layers. This sample yields a THG peak amplitude that is a factor of 5.8 (or 33 in peak intensity) larger than the one obtained with the single graphene sheet. We do not observe further increase of the THG peak amplitude using a stack of 10 or 15 layers due to linear losses experienced by the pump and the third harmonic signal. To monitor the level of inhomogeneity on each graphene-stack samples, data is collected on 5 different spatial spots separated by 700 µm. We find standard variations of the THG signal of 22% for the single graphene layer (error bars in Fig. 2c). Spatial fluctuations in nonlinear properties can be attributed to an inhomogeneous distribution of impurities, such as PMMA residues, and uneven sample growth quality leading to changes in the doping level and electronic transport properties[30]. Much smaller fluctuations of ~6% are measured with the 10- and 15-layer samples due to a smoothing effect of the spatial inhomogeneities across the different graphene layers. The displayed amplitudes in Figs. 2b and 2c correspond to the THG field after the graphene sample as we multiply the measured amplitude by 1.9 to account for the total experimental losses at 2.4 THz. Three effects leading to a signal attenuation are considered: (i) a 25% decrease of the THG signal is due to the gating pulse duration (108 fs FWHM), which reduces the detection sensitivity towards higher THz frequencies[31]; (ii) a 11% lower detection efficiency at 2.4 THz is due to phase matching conditions in the THz detection crystal (see Section 3 in the Supplementary Information); (iii) a 21% is lost because of the transmission properties of the HPF (see Section 2 in the Supplementary Information). Taking these factors into account, we found a field conversion efficiency of ~0.2% in 6-layer graphene. This value is twice the maximum efficiency previously reported for a single-layer graphene[23], which was obtained with an accelerator-based super-radiant source producing longer multicycle pulses at a lower fundamental frequency of 0.3 THz. Note that a longer pump pulse duration in our experiments would likely lead to larger optical nonlinearities induced by deposited heat[26]. Additionally, a lower driving frequency is expected to produce a stronger nonlinear THz conduction in graphene[22,32].

We perform theoretical calculations of the THG spectrum and peak amplitude based on the nonlinear wave equations[8] using the experimentally measured pump spectrum and peak driving field. Fig. 2a shows the calculated spectrum generated within a single graphene sheet (black dashed line) using a nonlinear coefficient $\chi^3 = 2.4 \times 10^{-10}$ m$^2$/V$^2$, which is obtained by fitting the THG peak value. To allow direct comparison between this calculation and raw experimental spectral curves in Fig. 2a, we divide the theoretical result by 3.6 (square of the factor 1.9 described above for the field amplitude) to account for experimental losses. The third-order nonlinear coefficient used in this calculation is slightly lower than the one reported in previous work in a similar experiment[27]. Discrepancies between our results and values previously reported can be attributed to the relatively short THz pump pulse duration in our experiments, potentially some residual water vapor absorption, or the fact that different graphene samples have slightly different properties due to fabrication and sample preparation[33]. In Fig. 2c, nonlinear calculations using the parameters found for a single-sheet graphene are performed for multilayered graphene samples from which we can extract the THG peak amplitude. We consider an 4% loss for both the THz pump and third harmonic amplitude as they transmit through each graphene layer (see Section 1 in the Supplementary Information). This loss in the THz region is consistent with values reported by previous work[30,32,34]. Furthermore, we also perform the same calculation with a third-order nonlinear coefficient that is a function of the field strength according to previous experimental observation in graphene[27] (see Section 1 in the Supplementary Information). We find an overall agreement between the model and experimental results, and no major differences when field-dependent nonlinearities of graphene are introduced in the model. The main discrepancies between experiment and calculations, which arise when the number of layers exceeds 3, can be in part attributed to variations of the nonlinear properties across different graphene sheets stacked together. Our calculations show that a maximum THG signal is expected when 9 graphene layers are used. This signal is, however, only 14% larger than the one calculated for 6 layers. As a result, the higher nonlinear conversion efficiency experimentally observed with a 6-layer graphene sample is likely an anomaly caused by sheet-to-sheet fluctuations of the nonlinear properties of graphene.

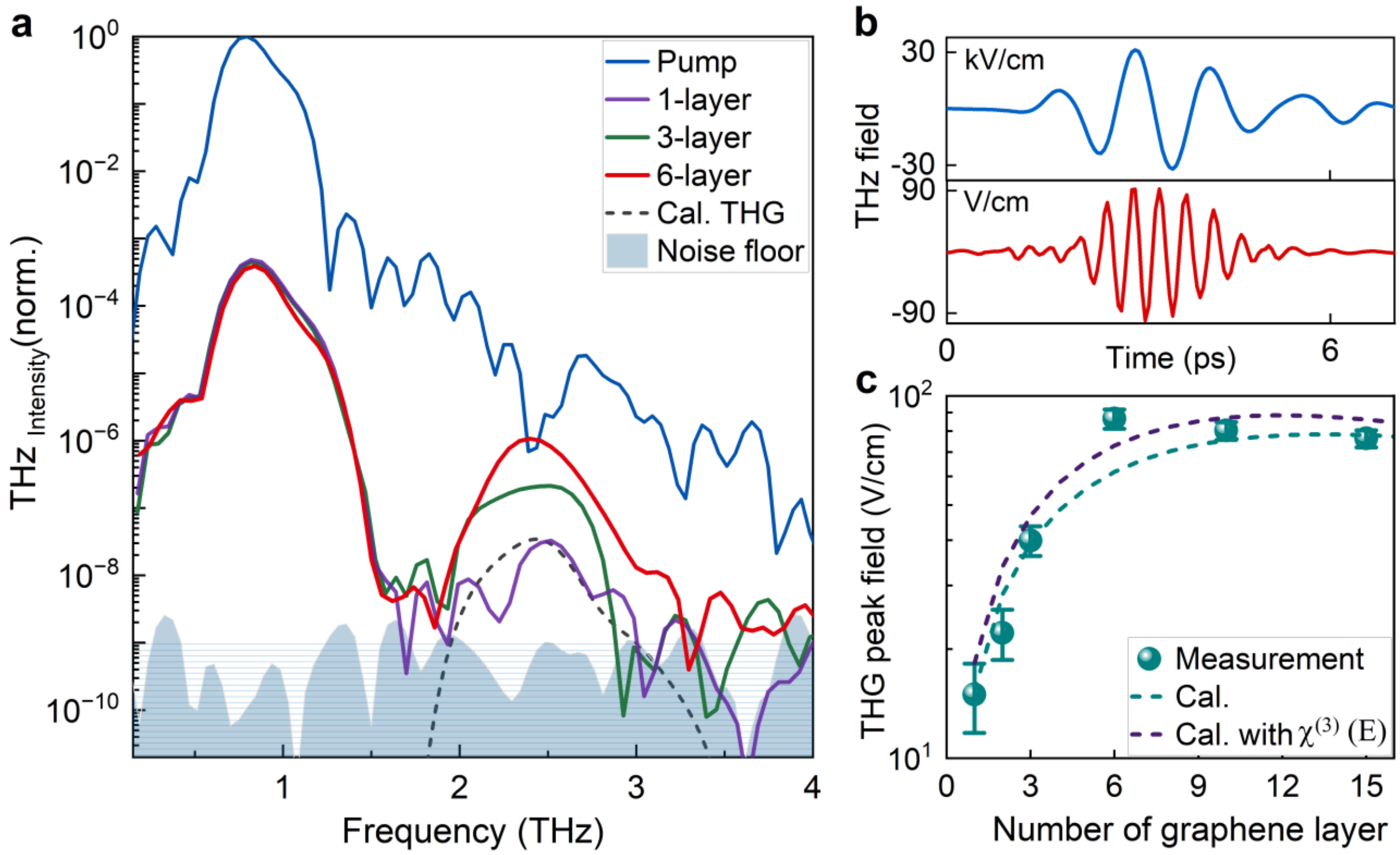

**Fig. 2 Measured THG in stacked graphene layers. a** The blue line represents the spectrum of normalized THz pump intensity transmitted through a LPF. A HPF is used to collect the other measurements. THG signature is observed around 2.4 THz for stacked graphene layers: 1-, 3-, and 6-layers. The grey shaded area shows the noise floor. The dashed black line is a calculation of the THG spectrum produced by a single-layer graphene also accounting for the experimental detection efficiency. **b** Time-domain THz transient of the pump pulse (blue line) and third harmonic pulse (red line) collected with a 6-layer graphene. The y-axis units for the fundamental pulse are kV/cm and they are V/cm for the third harmonic **c** THG peak field measured in samples with different numbers of graphene layers (green circles.) compared with calculations (dashed lines). The error bars correspond to the standard deviation calculated from five measurements on different spots.

## Electrical gating tunability of stacked graphene

Graphene's electronic THz nonlinearity has been attributed to the interaction of the THz field with free carriers and the subsequent interplay among electronic, phononic, and thermodynamic processes[6,22,35]. Therefore, a variation of the nonlinear response can be induced by an electrical gate changing the carrier density.[26] We use 1-, 2-, and 3-layer graphene samples electrically connected to a gate to simultaneously study the effect of carrier density and a multilayered design on nonlinear effects. The density is controlled using a polymer electrolyte (LiClO4:polyethylene oxide (PEO)) gate[36,37] connected to each graphene layer of the stack (see Methods). The electrical resistance of the samples is monitored by measuring the source-drain current as a function of the gate voltage. From these measurements, one can define the voltage offset $V_0$ required to reach the minimum free carrier density, also referred to as the charge neutrality point. As shown in Fig. 3a, the total resistance decreases as we increase the number of layers because they are connected in parallel (see Fig. S4 of the Supplementary Information). By scanning the gate voltage $V_{gate}$, we obtain a standard resistance response with a peak corresponding to the minimum doping concentration while the n- and p- doped region correspond to the negative and positive region of $V_{gate}$ -$V_0$, respectively. We vary the gate voltage and monitor the THG power calculated by integrating the spectrum centered at 2.4 THz over the $1/e^2$ bandwidth. Fig. 3b shows experimental measured THG power (circles) as a function of $V_{gate}$-$V_0$ where the lines correspond to b-spline interpolation to guide the eye. With a single graphene layer, we find a minimum THG power at the charge neutrality point because less carriers are contributing to the material nonlinearities[26]. We find two maxima symmetrically on each side of this point, at $|V_{gate} - V_0| = 0.7$ V. The ratio between the two extreme values is $\varepsilon_{1G} = 2.3$, which confirms that controlling the carrier density is essential for optimizing the THG efficiency. As we further increase both the n- and p-doping concentration, the nonlinear signal decreases by about 50%. This behavior agrees with previous reports on the emergence of a metallic phase at high doping concentration gradually reducing the material's thermodynamic nonlinearity[26].

Interestingly, the dependence of the THG power as a function of the gate voltage is different when we add more graphene layers. In the case of a 2-layer graphene sample, the dependence on $V_{gate}$ closely resembles the one observed with a single layer. However, we find $\varepsilon_{2G} = 1.9$, which is lower than the same ratio measured with a single layer $\varepsilon_{1G}$. For 3 layers, the signal as a function of $V_{gate}$ is more erratic, with a peak, instead of a dip, around $V_{gate}$-$V_0 = 0$, and what appears to be random oscillations as a function of $V_{gate}$. Nonetheless, we measure $\varepsilon_{3G} = 1.6$, indicating the significant effect the gate still has on the carrier density to enhance the sample's nonlinear effects. The lower values of $\varepsilon$ as the number of graphene layer is increased can be attributed to a different intrinsic doping concentrations across the layers, resulting in different values for $V_0$[33]. As a result, it is unlikely that the neutrality point of multiple superimposed graphene layers can be reached simultaneously with a single gate. Also, because of the device geometry, the change in carrier density may be different across the layers because of a screening effect, potentially causing larger variations inside the layers closer to the gate on top. Therefore, a multilayered architecture able to independently control carrier density in each graphene sheets would be optimal to reach maximum nonlinear effects. In this experiment, the 3-layer sample displays the highest nonlinear conversion efficiency due to its longer interaction length even though gate-induced enhancement of nonlinear effects is slightly lower as discussed above. Fig. 3c shows the minimum and maximum THG power obtained while changing $V_{gate}$ with the 1-, 2-, and 3-layer gated graphene samples. These two sets of data when expressed in their corresponding peak field values follow a linear dependence (dashed lines) with the number of layers. The significant separation between these lines attest to the importance in controlling graphene's carrier density to optimize nonlinear effects.

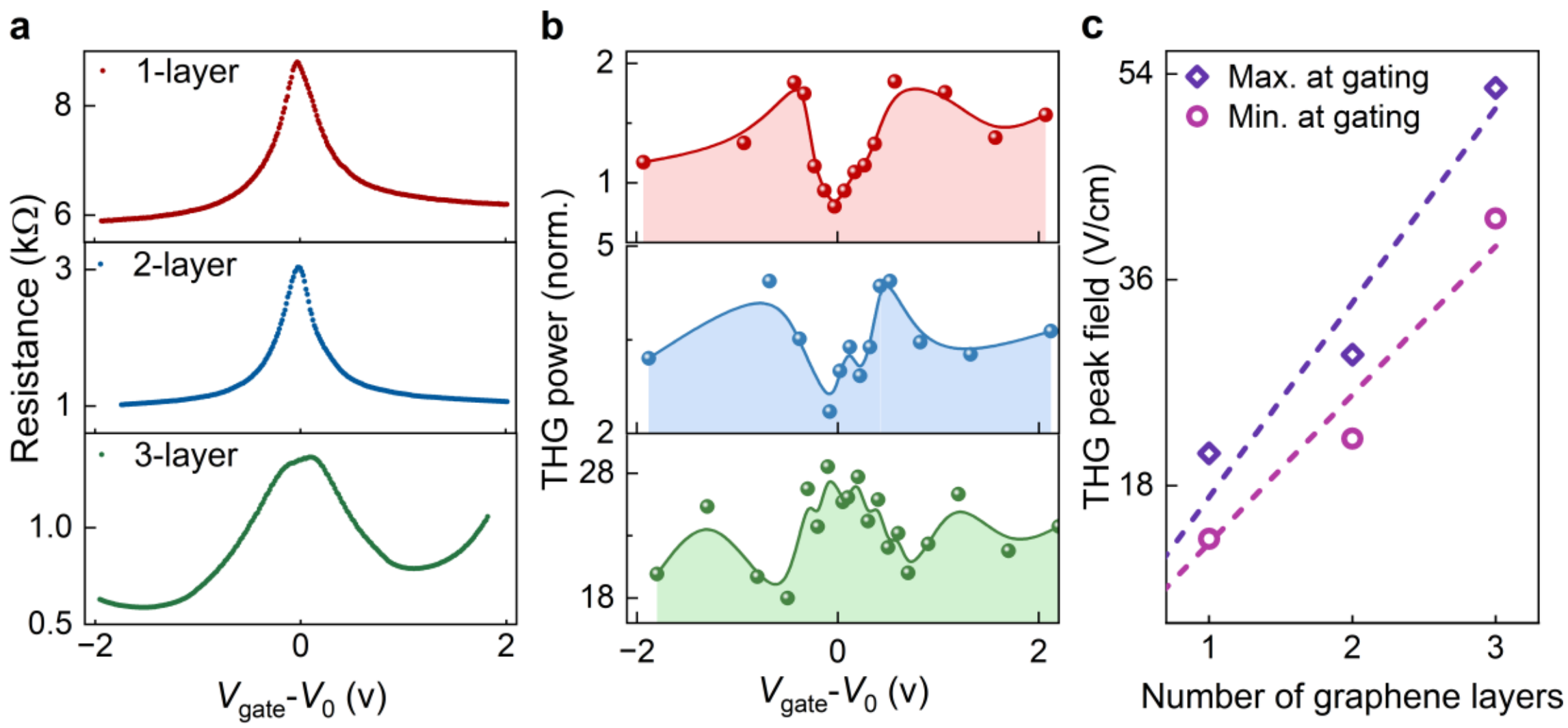

**Fig. 1 Electrical tunability of terahertz nonlinearity in gated multilayer graphene sheets. a** Experimentally measured sheet resistance and **b** extracted THG power for 1-layer (red), 2-layer (blue), and 3-layer (green) graphene sheets as we vary the gate voltage ($V_{gate}$). $V_0$ represents the applied voltage necessary to reach the maximum resistance of graphene, corresponding to the lowest carrier density. The circles represent measurements, and the lines depict a b-spline interpolation used as a guide to the eye. **c** Maximum and minimum THG peak field amplitudes measured on all gated multilayered samples while varying the gate voltage. The dashed line shows a linear fit.

## Metamaterial-graphene architectures

In a third series of experiments, we explore the effect of metasurfaces on THG in an architecture combining multilayer graphene and an electrical gate. Three different designs of metasurfaces are considered based on their distinctive spectral transmission properties at the THz pump frequency. Fig. 4a shows a schematic of these samples. They consist of a cross-slot bandpass filter (BPF), a cross-shaped bandstop filter (BSF), and a wire-grid polarizer (WGP). Two superimposed graphene sheets acting as the nonlinear medium are transferred on each metasurface. A transparent polymer electrolyte gate deposited on top of the structure is used to vary the carrier density (see section 4 of the Supplementary Information). Fig. 4c shows the measured THz transmission spectrum of each device with the fundamental resonance of the BPF and BSF centered at the THz pump frequency of 0.8 THz. We use FDTD simulations (Lumerical) to visualize the electric field distribution within a metasurface's unit cell at the height corresponding to the 2-layer graphene sample (Fig. 4b). We observe a significant enhancement of the incident THz field in the gap between metallic elements, locally reaching more than an order of magnitude. Since the third harmonic field amplitude $E_{3\omega}$ has a cubic dependence on the driving field $E_{\omega}$, this effect directly leads to a more efficient THG. Fig. 4d shows the measured third harmonic power normalized to the value obtained with the bare 2-layer graphene. We distinguish measurements collected when no gate voltage is applied ($V_{gate} = 0$) (yellow columns) and when $V_{gate}$ is optimized to reach the maximum THG power (blue columns). The BSF-graphene architecture leads to the largest THG power. We measure an enhancement factor of 2.5 when the sample is non-gated and a factor of 3 when $V_{gate}$ is optimized to the maximum THG power. The WGP-graphene architecture increases the THG power by a factor of 2 for non-gated approach, similar to the results presented in previous work for a similar device at the same peak field.[27] The use of BPF slightly reduces THG if the sample is not gated. Although this structure locally displays a high field enhancement, this feature is partially counteracted by a large fraction of the sample covered by the metallic structure, and therefore not contributing to nonlinear effects. For both BPF and WGO, the effect of gating can increase the THG signal by 20% and 75%, respectively.

To theoretically estimate the impact of the metasurface substrate on the nonlinear conversion efficiency, we use numerical simulations to calculate the field distribution and we then integrate $E_\omega^3$ within a metasurface's unit cell (uc). This value is compares to a reference obtained without the metasurface. In our calculations, we incorporated the induced field dependence of the third-order nonlinear coefficient, $\chi^{(3)}(E_\omega)$, using a rational fitted model derived from previously work[27] (see Section 1 in the Supplementary Information for simulation details and fitting model). The ratio $\gamma = \oint_{uc} \chi^{(3)}\, E_{\omega,MS}^3\, da / \oint_{uc} \chi^{(3)}\, E_{\omega,0}^3\, da$ indicates the THG enhancement factor induced by the metasurfaces. In Fig. 4d, we compare $\gamma$ (magenta columns) to the corresponding experimental results. We find a good agreement between the calculated and measured THG power. The slight discrepancies may arise from variations in the enhancement of plasmonic structures between simulations and fabricated samples, particularly at the sharp edges. Additionally, minor shifts in resonance may occur due to deviations in dimensions during the fabrication process.

Numerical simulations reveal that the field amplitude at the driving frequency remains nearly constant along the direction normal to the substrate. Within a 2 μm distance, the amplitude variation remains below 5%. This ensures that similar metasurface-induced enhancement factors of the THG signal will be observed regardless of the number of decoupled graphene layers in the sample. This remains true for thick samples since, after 15 graphene layers, additional layers contribute less to the THG signal due to absorption of the driving field. Therefore, considering a sample with 6 to 9 decoupled graphene sheets, electrically connected to allow independent control of carrier concentration in each layer, and deposited on a BSF metasurface substrate, we can envision a total THG enhancement factor exceeding 100.

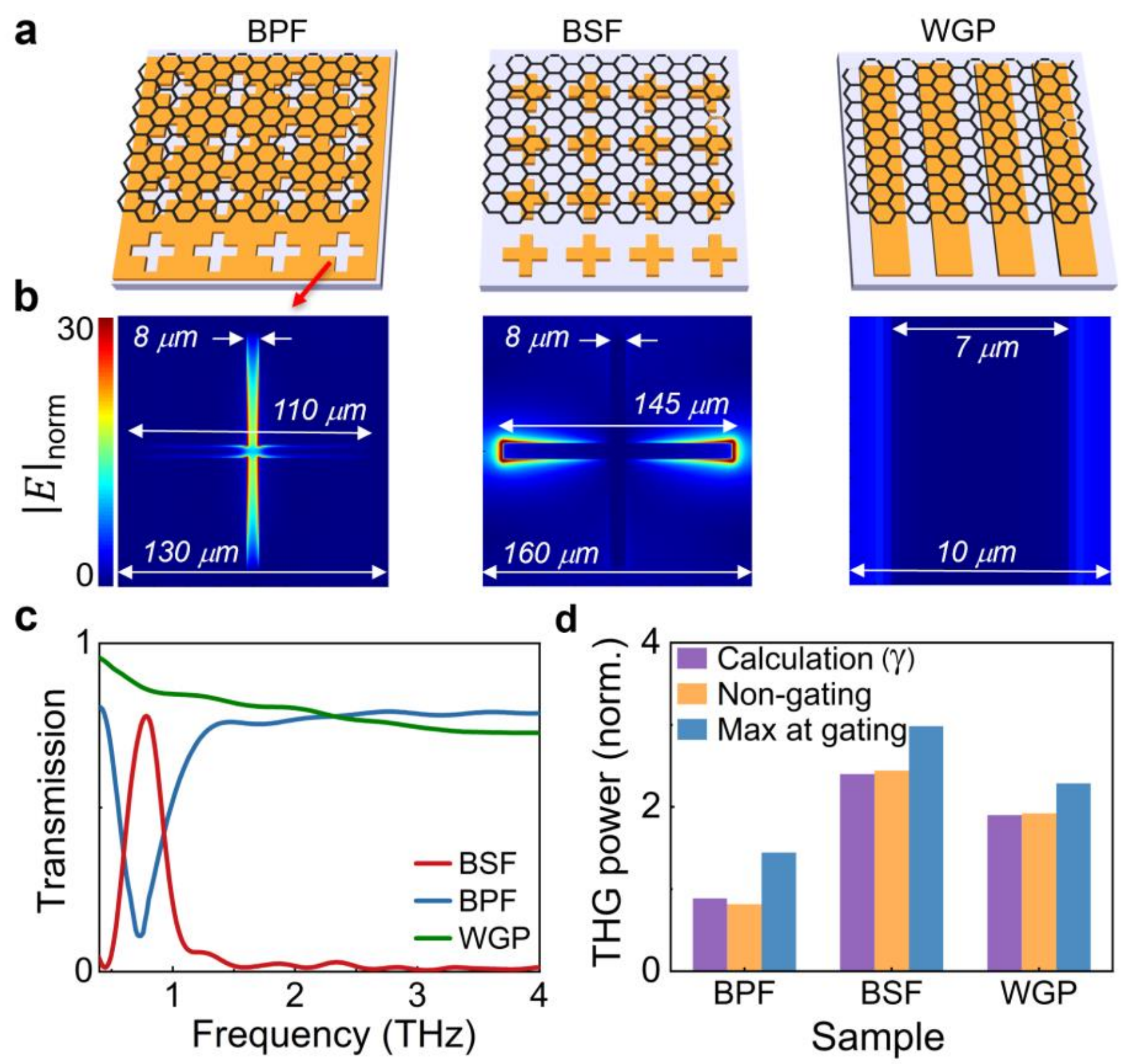

**Fig. 2 Metamaterial-graphene-based architectures**. **a** Schematic of a cross-slot bandpass filter (BPF), cross-shaped bandstop filter (BSF), and a wire-grid polarizer (WGP) with multilayer graphene deposited on their top surface. **b** FDTD simulated electric field distribution inside graphene for the three architectures. Metallic elements are shown in dark blue. **c** Measured THz power transmission of the metamaterial structures displaying fundamental resonances around 0.8 THz for the BPF and BSF. **d** Metasurface-induced enhancement of the THG signal. We show numerical calculations $\gamma$ compared to experimental THG power normalized to the value obtained with a bare 2-layer graphene. We distinguish measurements collected when no gate voltage is applied ($V_{gate} = 0$) and when $V_{gate}$ is optimized to obtain a maximum THG.

## Discussion

We investigate third harmonic generation (THG) in graphene in the THz range while relying on three types of device architectures to make this nonlinear process more efficient. We experimentally investigate nonlinear samples with (i) a different number of graphene layers to increase the nonlinear interaction length, (ii) an electrical bias with a gate voltage to optimize the carrier density, and (iii) metallic metasurface substrates to locally enhance the pump field. Especially, we explore the potential to combine these techniques together within one structure. The largest improvement of the THG signal is observed by increasing the number of graphene layers. In particular, the most efficient THG is measured with a 6-layered graphene sample, representing a trade-off between the nonlinear interaction length and linear absorption, which leads to a third harmonic power 33 times higher than obtained with a single graphene layer. We then use a simple configuration allowing a transparent electrical gate to tune nonlinear properties of graphene-based samples. In a gated single graphene layer, THG can increase by a factor of 2.3. Although this factor is lower when the sample contains multiple graphene layers, the gating voltage always leads to a significant increase of the THG signal, larger than 60% in our experiment. Plus, a design allowing independent tuning of the free carrier concentration in each layers could be used to ensure a two-fold THG enhancement in multilayered designs. In a third series of experiments, different types of metasurfaces are used as substrate to explore the effect of local field enhancement on THG in graphene. Interestingly, the most efficient design is a bandstop filter (BSF) at the fundamental frequency, featuring a 3-fold increase in THG power. We propose a model based on numerical calculations to estimate the nonlinear signal enhancement induced by a metasurface, which provide good agreement with experimental results. More importantly, our work demonstrates the possibility to combine these strategies to potentially enhance THG by more than two orders of magnitude. The sample architectures can also apply to other types of 2D nonlinear materials, such as transition-metal dichalcogenides[38], and they can be combined with other techniques to enhance nonlinear effects, such as the use of topological insulating substrates[13,39]. Finally, our study highlights the potential of multilayered graphene with electrical gating and metamaterial substrate to achieve unprecedented nonlinear responses, thus offering promising avenues for engineering graphene-based nonlinear devices. Such advancements are essential to enable efficient chip-integrated nonlinear THz signal converters that will be used in future communication technologies.

## Materials and methods

### Sample preparation

We use commercial monolayer CVD-grown graphene on copper foil, coated with a poly(methyl methacrylate) (PMMA) layer, obtained from Graphenea. Graphene layers of 1 cm × 1 cm dimensions are wet-transferred onto a 188 µm thick Zeonor substrate[30] (a cyclo-olefin copolymer transparent to THz radiation with a refractive index of n ~ 1.53 at 1 THz). This is followed by the deposition of a 60 nm-thick $Al_2O_3$ layer using an electron-beam evaporator (NexdepSeries,

Angstrom). To form multilayer graphene samples, additional graphene layers and alternating $Al_2O_3$ layers are stacked on the same substrate.

The gating voltage is applied with a 5nm Ti/20nm Pd/150nm Au electrodes deposited on the sides of the graphene layers using a single evaporation system and a shadow mask placed atop the graphene surface. The source and drain contacts make direct contact with the graphene layer, while the gating contacts has no direct contact. A spray-coating technique is employed to apply a 400 nm-thick transparent polymer electrolyte onto the graphene layers. The electrolyte is made of Polyethylene oxide (PEO) and LiClO4 in an 8:1 weight ratio, which is dissolved in a methanol solution. Subsequently, the sample is affixed onto a custom-printed circuit board (PCB) with a central hole to perform optical transmission measurements. The electrical measurements are performed at constant voltage while monitoring the current with Keithley source meters (models 2400).

The metasurface are fabricated on a 188 µm thick Zeonor substrate using a conventional positive photolithography process. Aluminum is deposited onto the patterned substrate using a sputtering technique, followed by a lift-off process to achieve 200 nm-thick metallic arrays. Subsequently, a 60 nm-thick $Al_2O_3$ layer was deposited on the metasurfaces using an electron-beam evaporator, after which graphene sheets were wet-transferred onto them.

**Experiment**

In our experimental setup, high-field (360 kV/cm) single-cycle THz pulses are generated using the tilted pulse front technique in a $LiNbO_3$ crystal. These pulses originated from an amplified 10 kHz Yb-laser system with following parameters: central wavelength of 1030 nm, a pulse width of 170 fs, and a pulse energy of 1 mJ. Detection involved electro-optical sampling method by utilizing pulses from an optical parametric amplifier (OPA) at 960 nm wavelength, with a pulse duration of 108 fs (FWHM), and employing a 1 mm-thick GaP nonlinear detection crystal in an electro-optic sampling scheme. The gating pulses at 960 nm satisfy phase matching condition in GaP to achieve high detection sensitivity at the third harmonic frequencies at 2.4 THz. All experiments are conducted at room temperature inside a dry-air purged enclosure.

**Simulations**

Numerical simulations are performed with a 3D finite-difference time-domain (FDTD) solver (Lumerical Inc.) to design the plasmonic metasurface resonators with fundamental resonances around 0.8 THz and explore the electric field distribution of the THz pump pulse on the structures. We utilize periodic boundary conditions along the in-plane axis. Additionally, a perfectly matched layer (PML) is applied in the direction of optical propagation to absorb all incident fields, effectively preventing any reflections at that interface. For the simulation of electric field spatial distributions, we employed a minimum override mesh size of 50 nm along the surface of the structures and 4 nm along the THz propagation direction.

**Acknowledgements**

We acknowledge funding from the Natural Sciences and Engineering Research Council of Canada (NSERC) Discovery funding program (RGPIN-2023-05365) and the University of Bayreuth Centre of International Excellence “Alexander von Humboldt”. A. Maleki, G. Herink and J.-M. Ménard acknowledge financial support from the Mitacs Globalink Research Award. We thank B. Sullivan, A. Jaber, S. Ntais and S. Biberger for insightful discussions and technical support.

**Author Contributions**

A.M., M.B.H., G.H., and J.-M.M. conceptualized the idea and designed the experiment. A.M. completed the FDTD simulations for the metasurfaces and Y.X. fabricated the metasurfaces. A.M. engineered the gated and non-gated multilayer graphene samples, and spectral filters. A.M. and M.B.H. carried out the measurements. A.M. and J.-M.M., analyzed the experimental results. R.W.B., G.H., and J.-M.M. supervised this work.

**Conflict of interest**
The authors declare no competing interests.

**Data availability**
The experimental data that support the findings of this work is available upon reasonable request from the corresponding authors.

**Supplementary information**
Included.

## Supplementary information

# Strategies to enhance THz harmonic generation combining multilayered, gated, and metamaterial-based architectures

Ali Maleki,[1] Moritz B. Heindl,[2] Yongbao Xin,[3] Robert W. Boyd,[1,4,5] Georg Herink,[2] Jean-Michel Ménard[1,4,*]

[1]*Department of Physics, University of Ottawa, Ottawa, ON K1N 6N5, Canada*
[2]*Experimental Physics VIII – Ultrafast Dynamics, University of Bayreuth, Bayreuth, 95447, Germany*
[3]*Iridian Spectral Technologies Ltd, Ottawa, ON K1G 6R8, Canada*
[4]*School of Electrical Engineering and Computer Science, University of Ottawa, Ottawa, ON K1N 6N5, Canada*
[5]*Institute of Optics and Department of Physics and Astronomy, University of Rochester, NY 14627, USA*

**Email: Jean-Michel Ménard (jean-michel.menard@uottawa.ca)*

### Section S1: Calculations of the third harmonic generation (THG)

Fig. S1a depicts the schematic of a multilayer graphene nonlinear sample, illustrating the interaction of a pump pulse with graphene layers and resulting THG signals on each layer. The THG signals produce in each graphene sheets accumulate as the THG pulse propagates in the structure. In detail, the THz pump pulse ($E_\omega$) first impinges on the Zeonor substrate, experiencing an 8% power transmission loss due to Fresnel reflections at the air-substrate interface. Then, successive third harmonic fields ($E_{3\omega}^i$) are produced by the THz driving field ($E_\omega^i$) incident on each graphene layers, where $i$ = 1, 2, 3, … refers to the position of the graphene sheet in the multiplayer stack. During the THz pulse propagation through a stack of graphene sheets, we consider that 4% of the pump pulse amplitude is absorbed in the first graphene layer. The remaining pump pulse then proceeds to interact with the second layer to generate a slightly weaker THG ($E_{3\omega}^2$) and where another 4% gets absorbed. This absorption and signal generation process continues for subsequent graphene layers. It is important to note that we also consider a 4% absorption of the third harmonic field amplitude as this component propagates through each graphene layer. Considering both the absorption of the pump pulse ($\alpha_\omega$) and third harmonic components ($\alpha_{3\omega}$), we establish a relationship for THG in n-layer graphene structures as follows:

$$E_{3\omega}^{Total} = \sum_{i=1}^{n} E_{3\omega}^{i} = \frac{3\omega d}{8cn_{3\omega}} \chi^{(3)} \sum_{i=1}^{n} \left[(1-\alpha_\omega)^{3(i-1)}(1-\alpha_{3\omega})^{n-i} E_{3\omega}^{1}\right],$$

where $c$ is the speed of light, $d$ = 0.3 nm is the thickness of each graphene layer, $n_{3\omega}$ = 10 is the refractive index of the graphene layer, $\omega$ is the pump radial frequency, and $\chi^{(3)}$ is the third order susceptibility. For these calculations, we used the linear properties of graphene reported in previous work[1] and nonlinear coefficient of $\chi^3 = 2.4 \times 10^{-10}$ m$^2$/V$^2$ obtained from our THG data obtained with a single graphene layer.

Since the coefficient $\chi^{(3)}$ of graphene has been observed to be a function of the driving field amplitude, we extract this dependence from previously reported data[1] (circles in Fig. S1b). We calculate a fitting function: $\chi^{(3)} = \frac{7.2e^{-10}}{1+1e^{-9}E_{\omega}+12e^{-4}{E_{\omega}}^{2}}$ in m²/V², where $E_{\omega}$ is the driving electrical field amplitude in kV/cm. This model considers a maximum value for $\chi^{(3)}$ as suggested by the thermodynamic calculations of Ref. [1].

This expression for $\chi^{(3)}$ **is not only used to estimate the THG in a multilayer graphene structure (purple line in Fig. 2c), but also determine the effect of an inhomogeneous field distribution induced by a metasurface substrate on the THG (purple column in** Fig. 4d). We define the metasurface-induced enhancement factor as:

$$\gamma = \oint_{uc} \chi^{(3)} \left(E_{\omega,MS}\right)^{3} \, da / \oint_{uc} \chi^{(3)} (E_{\omega})^{3} \, da,$$

which uses both the calculated electric field spatial distribution inside a graphene layer above the metasurface ($E_{\omega,MS}$), and a homogeneous field impinging on a graphene layer with no metasurface substrate ($E_{\omega}$). The field modulation induced by the metasurface is simulated using a mesh size of 50 nm along the surface of the structure and 40 nm along the optical propagation direction. Decreasing the mesh size does not alter the results of the calculations. However, increasing the mesh size by a factor of 2 in the in-plane direction, to reach lower spatial resolution, decreases the calculated factor $\gamma$ by about 5%.

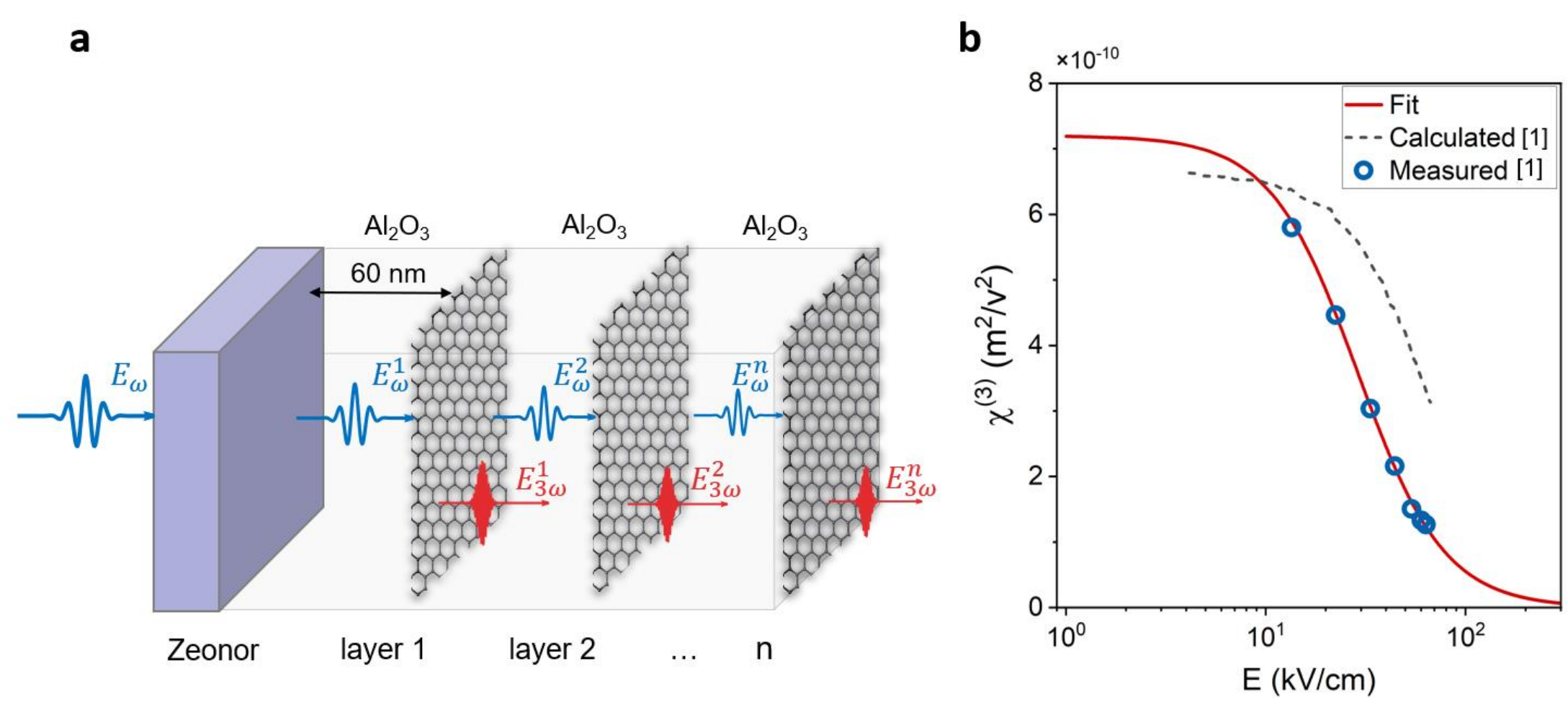

**Fig. S3 a** Schematic of the THz propagation in multilayer graphene sheets. THz pump pulses on each graphene layer are shown with blue pulses, and THG pulses are shown in red. The schematic of the nonlinear sample comprises a substrate (Zeonor in our experiment) and graphene layers (n-layer), that are separated by $Al_2O_3$. **b** Measured (circles) and calculated (dashed line) third order nonlinear coefficients (**$\chi^{(3)}$**) based on the thermodynamic model for a single layer graphene reproduced from J-C. Deinert et al.[1]. Solid line in red shows our fitting function.

## Section S2: Lowpass filter (LPF) and highpass filter (HPF)

The schematic of the LPF and HPF structures and their corresponding intensity spectral transmission are shown in Fig. S2.

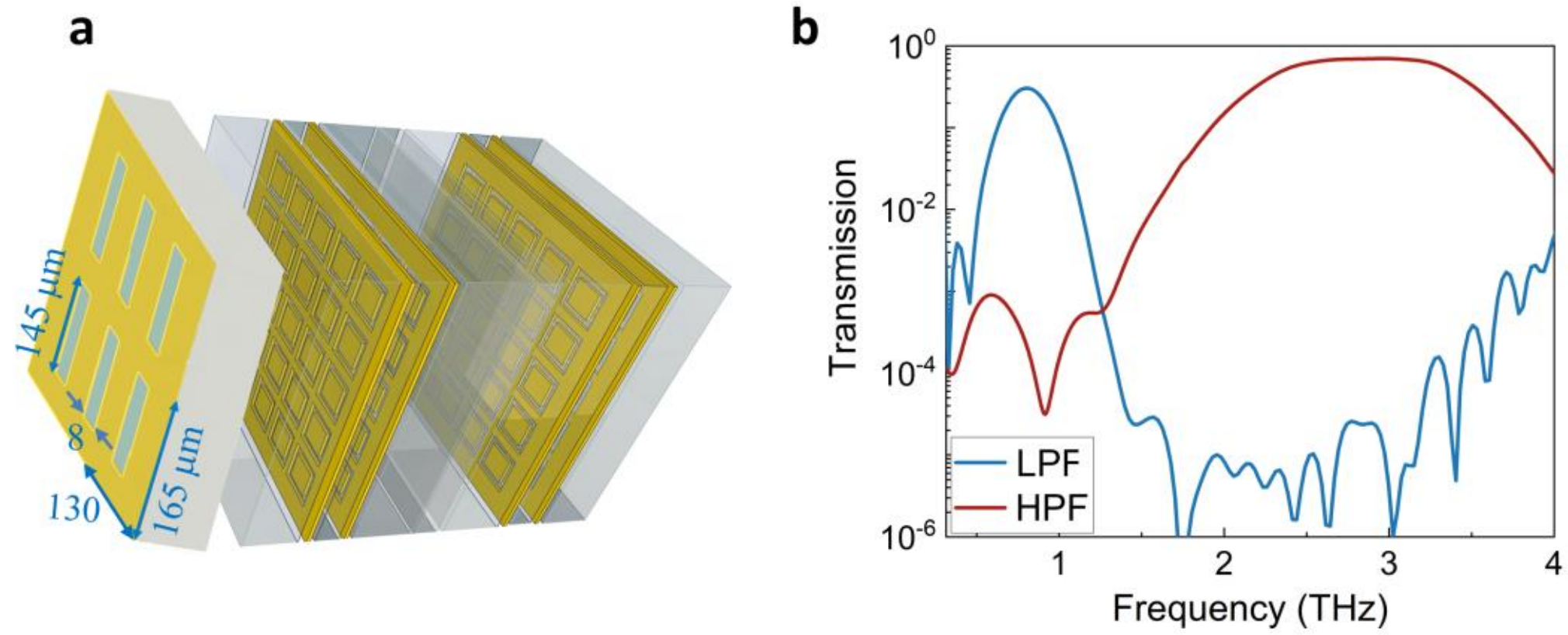

**Fig. S4 a** Schematic of the spectral filters. The highpass filter (HPF) is made of 4-layer square-slot plasmonic metasurfaces as described in our previous work[2]. During fabrication of the lowpass filter (LPF), we added a bar-slot bandpass filter on top of the 4-layer plasmonic structures to create a sharper bandpass around 0.8 THz and strongly attenuate any residual signal around 2.4 THz. **b** Intensity transmission of the LPF and HPF filters in the THz region.

## Section S3: Sensitivity of the GaP detection crystal

We use an electro-optical sampling method to monitor the THz spectrum. The gating pulse, produced by an optical parametric amplifier (OPA), is centered at 960 nm wavelength and has a pulse duration of 108 fs (FWHM) as shown in Fig. S3a. The autocorrelation is obtained by two-photon absorption in a Si photodiode. The deviation from the expected peak-to-background ratio of 8:1 is attributed to a small amount of three-photon-absorption. The wavelength is selected to optimize phase matching conditions at 2.4 THz in our THz detection crystal, which is a 1 mm-thick 110-oriented GaP crystal. The calculated detection efficiency at 2.4 THz is however still 11% lower than the detection efficiency at the pump pulse frequency of 0.8 THz. These calculations use the linear properties of GaP in the THz region[3] are shown in Fig. S3b. The factor of 11% is considered when we compare the field amplitude of the third harmonic and fundamental components.

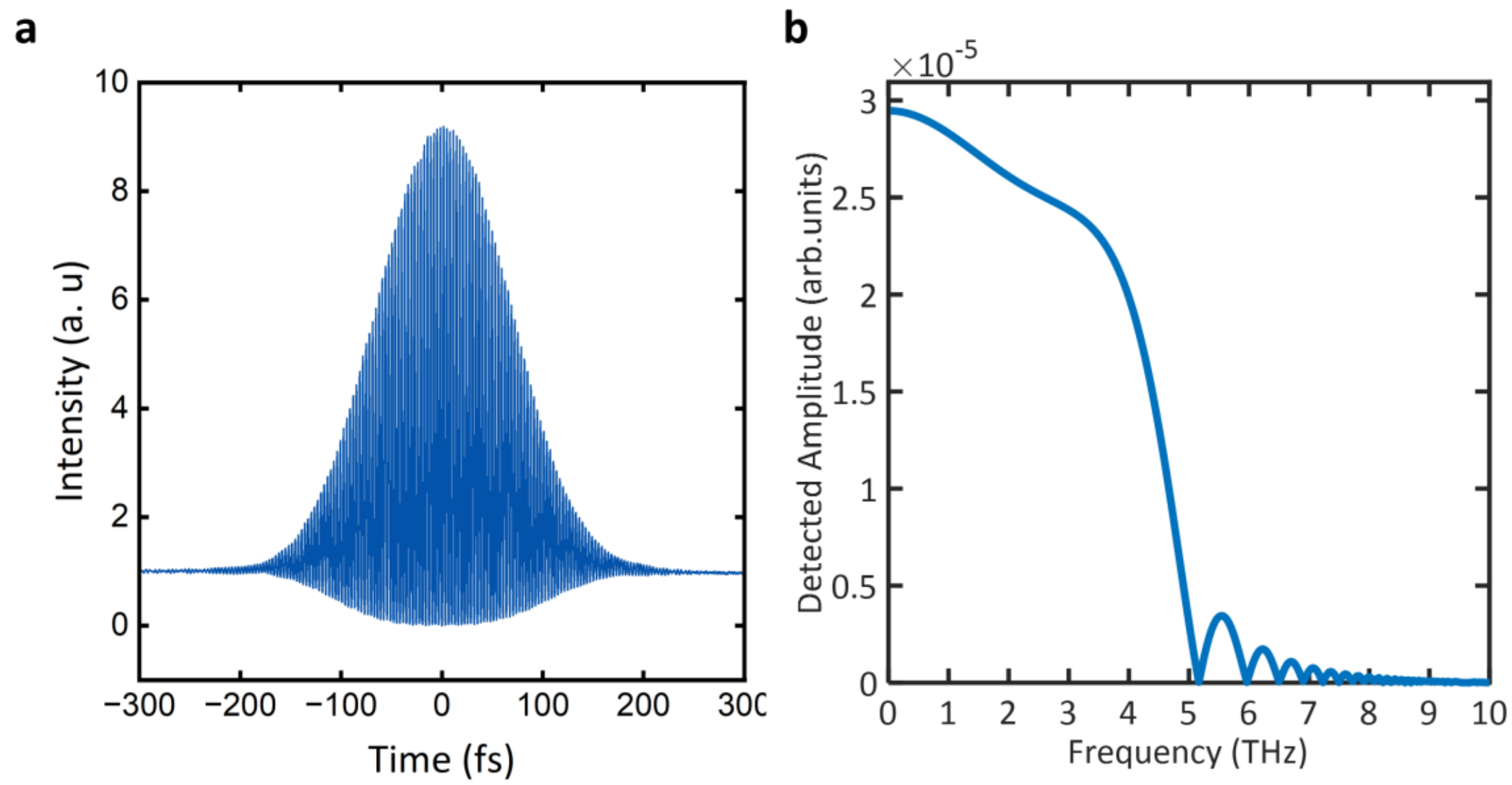

**Fig. S5 a** Measured interferometric autocorrelation of the gating pulse, representing intensity pulse duration of 108 fs at FWHM. **b** Simulated detection sensitivity of the GaP crystal at the gating 960 nm wavelength.

## Section S4: Fabrication of multilayer graphene architectures

We use monolayer CVD graphene on copper foil coated with a poly(methyl methacrylate) (PMMA) layer, and wet transferred them on substrates. This is followed by the deposition of a 60 nm-thick $Al_2O_3$ layer in between graphene layers. The CVD graphene layer initially exhibited p-type doping, typically originating from chemical residues in the preparation process, wet-transferring processes, interactions with the substrate, and deposited polymer electrolyte[4]. In the next step, a pair of 5 nm Ti/20 nm Pd/150 nm Au electrodes were deposited on the sides of the graphene layers, serving as source and drain terminals for electrical characterization. The graphene sheets were connected in parallel on the sides of the metallic electrodes, as shown schematically in Fig. S4. For each graphene sample, we introduced a gate offset $V_0$ to effectively compensate for the intrinsically hole-doping effect and shift the Fermi energy towards a neutrally charged Dirac point. This $V_0$ voltage varied from sample to sample, typically ranging between 0.1 to 0.3 V.

We use the same fabrication process to transfer graphene sheets on plasmonic metasurfaces. 60 nm-thick $Al_2O_3$ layers serve as spacing between graphene layers and the plasmonic structures to avoid the possibility of any graphene-metal interactions. To apply electrical gating on the graphene-metasurface samples, we employ a spray-coating technique to deposit 400 nm-thick transparent polymer electrolyte (as described in Methods) onto the graphene layers. Finally, in the high-harmonic generation experiment, THz pump light illuminates the structure from the substrate side, initially reaching the plasmonic side, and then interacting with the graphene sheets to achieve THG.

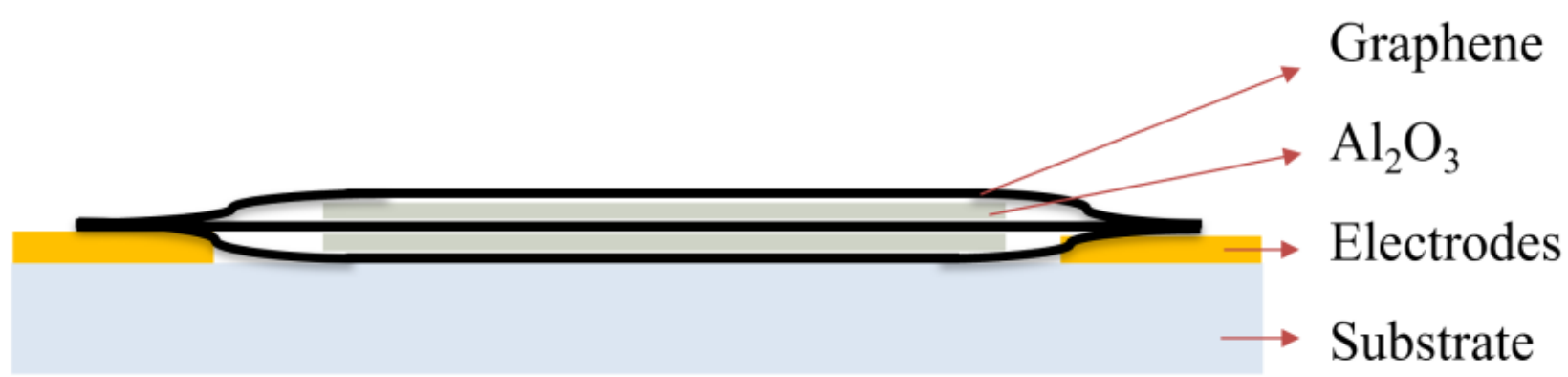

**Fig. S4** Schematic of the side-view configuration depicting graphene layers electrically connected in parallel and all attached to metallic electrodes.